\documentclass[12pt,a4paper,final]{iopart}

\newcommand{\revision}{\textcolor{black}}
\newcommand{\revisiontwo}{\textcolor{black}}

\usepackage[breaklinks=true,colorlinks=true,linkcolor=blue,urlcolor=blue,citecolor=blue]{hyperref}
\usepackage{iopams} 
\usepackage{graphicx}
\usepackage{color}
\usepackage[normalem]{ulem}

\begin{document}

\title{Group chasing tactics: how to catch a faster prey?}

\author[cor1]{Mil\'an Janosov}
\address{E\"{o}tv\"{o}s Lor\'and University}
\ead{janosovm@gmail.com}

\author{Csaba Vir\'agh}
\address{E\"{o}tv\"{o}s Lor\'and University}
\ead{viraghcs@hal.elte.hu}

\author{G\'abor V\'as\'arhelyi}
\address{MTA-ELTE Statistical and Biological Physics Research Group}
\ead{vasarhelyi@hal.elte.hu}

\author{Tam\'as Vicsek}
\address{Department of Biological Physics, Eötvös Loránd University \\ MTA-ELTE Statistical and Biological Physics Research Group}
\ead{vicsek@hal.elte.hu}


\begin{abstract}
We propose a bio-inspired, agent-based approach to describe the natural phenomenon of group chasing in both two and three dimensions. Using a set of local interaction rules we created a continuous-space and discrete-time model with time delay, external noise and limited acceleration. We implemented a unique collective chasing strategy, optimized its parameters and studied its properties when chasing a much faster, erratic escaper. We show that collective chasing strategies can significantly enhance the chasers' success rate. Our realistic approach handles group chasing within closed, soft boundaries -- contrasting most of those published in the literature with periodic ones -- and resembles several properties of pursuits observed in nature, such as the emergent encircling or the escaper's zigzag motion. 
\end{abstract}

\submitto{New J. Phys.}


\section{Introduction}

The animal kingdom is full of fascinating phenomena on every level. An essential part of this diverse biosphere is the complex biological interaction network between the different and within the same species in which interactions are categorized based on the benefits and harms of the interacting partners. An important type of these driving forces is the prey-predator interaction, which is also a very exciting example of collective behaviour in nature \cite{vicsek2012collective}.

Natural prey-predator systems have already been studied many times, including long term observations of population dynamics and a large number of field studies about animals' typical behaviour. One of the governing factors of these tendencies is the group hunting of large carnivores, like the lions of the Serengeti, the wild chimpanzees in the Tai National Park, the coyotes in the Yellowstone National park \cite{schaller1972serengeti, boesch1989hunting, gese2001territorial}. The results of these investigations revealed many interesting behavioural patterns of the animals, e.g. special hunting tactics and the size of the packs. 

Prey-predator systems have also raised the interest of theorists, who have constructed models to describe the behaviour of competing populations since Lotka and Volterra. As the field developed, the commonly used model schemes shifted from the continuous, differential equation based ones towards discrete models due to the advancements of computational technology \cite{pekalski2004short, antal2001critical, rozenfeld1999study}. In the past few years, agent based modelling also became popular using the concept of self propelled particles in the spirit of the model by Vicsek et al \cite{vicsek1995novel}.

In this work we propose a bio-inspired, continuous-space and discrete-time agent-based model of group chasing for the rarely studied scenario in which the evader is significantly faster than the pursuers. This realistic chasing approach overcomes many weaknesses of the previously published models. It combines many environmental factors like time delay, finite acceleration (inertia), external noise and closed boundaries (while most of the models in the literature are still using periodic boundary conditions). Moreover, it includes special collective chasing tactics, the chaser's prediction of their target's future position and the erratic escapers tactics as well \cite{angelani2012collective, lee2006predator, kamimura2010group, jin2010pursuit, saito2016group, ohira2015mathematical}. Significantly, our model operates universally in both two and three dimensions.

\subsection{Ethological background - chase and escape in nature}

\subsubsection{Chasing in nature}

Group chasing is one of the most important forms of collective behaviour. Its significance arises from the fact that cooperation among predators can notably increase their chances of catching even hard-to-catch prey which makes these collaborative species much more successful predators. Packs in which the cooperation is more efficient have better chances of survival, which puts an evolutionary pressure on the animals to optimize their strategies \cite{packer1988evolution, boesch1994cooperative, mech2010wolves, kruuk1972spotted}

Previous field studies have reported approximately 50 different species in which increases in group size above a certain point reduces their success rate per capita \cite{packer1988evolution}. A biologically logical explanation is that the group became too large and not all the predators were able to feed even if the hunt was successful, therefore the size of chaser groups generally do not  grow past a specific threshold \cite{stander1993hunting}. 

\begin{figure}
\begin{center}
\includegraphics[width=0.85\textwidth]{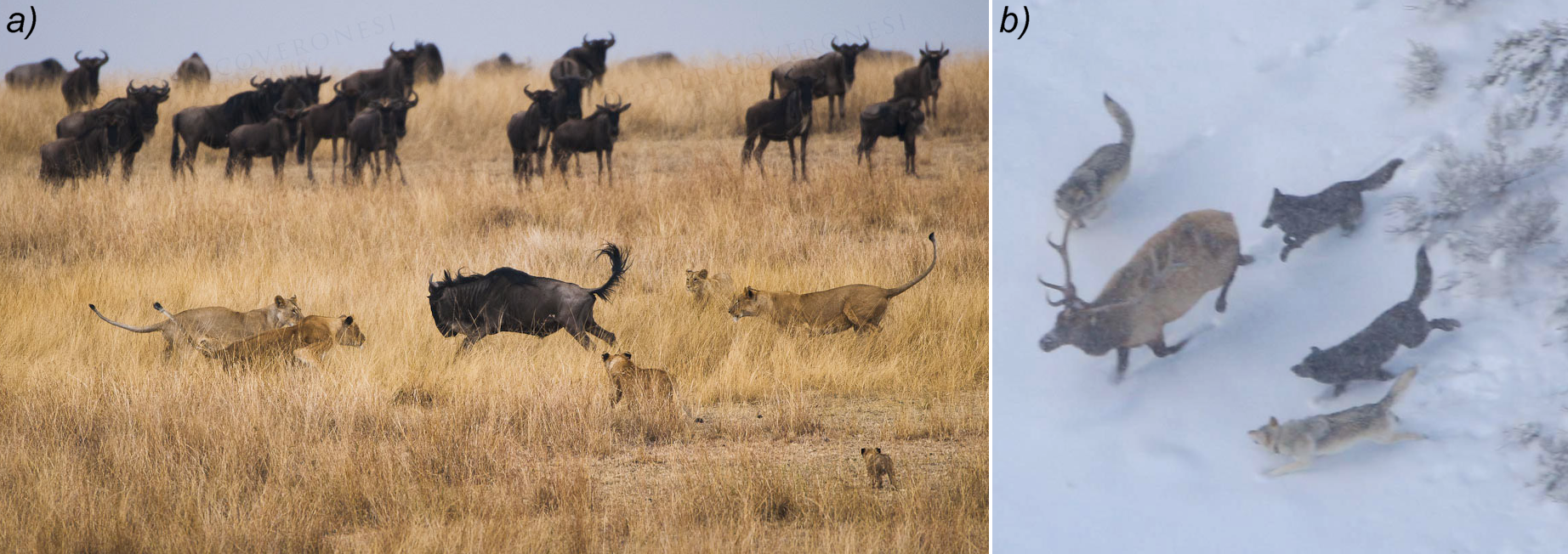}
\caption{{\it a)} Lions encircling a separated wild beast \cite{fig11}. {\it b)} Wolves closing on an elk \cite{fig12}.}
\label{fig:encirc}
\end{center}
\vspace{1pt}
\end{figure} 

The variety of predator--prey systems is extraordinarily rich within birds, mammals, fishes or even insects \cite{mech2010wolves, lopez2006bottlenose, boesch1989hunting, bednarz1988cooperative, eaton1970hunting, kim2005cooperative}. A surprising finding is that even cheetahs, the fastest animals in the world, show an affinity for hunting in pairs \cite{eaton1970hunting, packer1988evolution}. Additionally, wolves tend to hunt elk, while coyotes have been studied hunting for pronghorns in e.g. migrating corridors using landmarks as a strategic tool (reported in Yellowstone National Park). In both cases the prey is faster than the predator \cite{mech2010wolves, macnulty2014influence}. We can also conclude that even wolves, one of the most unwavering examples in the animal kingdom, give up their pursuit after roughly one-two kilometres (which hardly takes two minutes at their top speed).

\begin{figure}[b]
\begin{center}
\includegraphics[width=0.5\textwidth]{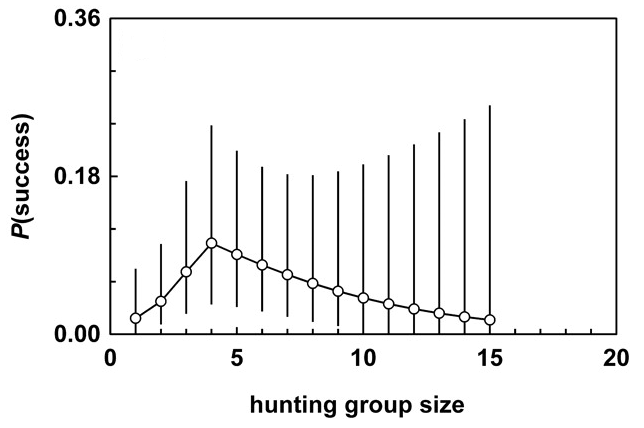}
\caption{The overall probability of success for a certain hunting pack of wolves -- elk encounter, depending on the size of the pack \cite{macnulty2011nonlinear}.}
\label{fig:optimalgroupsize}
\end{center}
\end{figure} 

The cooperation between hunters has different levels. For example, group hunting within bats just emerges completely unintentionally from the similar hunting methods of the individuals, while the study of large carnivores led to the finding that certain predators while hazing their prey tend to surround it collectively \cite{bailey2013group, brosnan2010cooperation}. This so called encircling was reported mostly regarding wolves and lions (Figure \ref{fig:encirc}), but even bottle-nose dolphins have been observed while encircling their prey \cite{lopez2006bottlenose, stander1992cooperative, packer1988evolution}.

Despite the limited available data, it is clear that for many predator species group hunting proved to be more beneficial then solitary hunting, even if not every member of the group participate in the pursuit. Evolution seems to have optimized the size of these hunting packs, because in the majority of cases, hunting pack size falls within the range of 3-10 (Figure \ref{fig:optimalgroupsize}).   

\subsubsection{Natural escaping strategies}

Fleeing is the most common response from animals against their chasers. First intuition would suggest that the only existing escaping strategy is to rush away from the chaser(s) as fast as possible. However, in real biological systems surprisingly different escaping strategies can also occur. Based on a concept proposed by Domenici et.al. we can differentiate between direct escaping and erratic escaping as separate strategies, although other kinds of strange actions like freezing or attacking back can also happen \cite{domenici1997escape, edut2004protean, humphries1970protean}. 

\begin{figure}
\begin{center}
\includegraphics[width=0.9\textwidth]{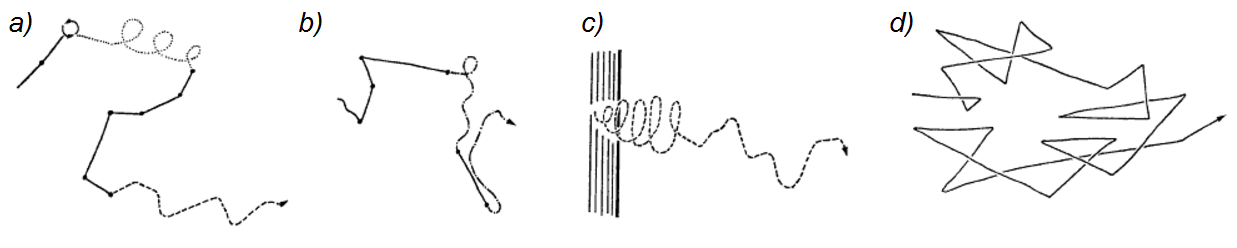}
\caption{Examples of erratic escaping -- combinations of  looping, whirling,  spinning and  zigzagging \cite{humphries1970protean}.}
\label{fig:zigzag}
\end{center}
\end{figure} 

Direct escaping means that the prey runs straight in the opposite direction from its chaser(s) and relies on its speed and endurance to survive. In other words, under ideal conditions the escaper can outrun its chaser(s), but the prey can be caught if it is slower, if it encounters unexpected landmarks or if the chasers strategies are more advanced. 

\begin{figure}[b]
\begin{center}
\includegraphics[width=1.0\textwidth]{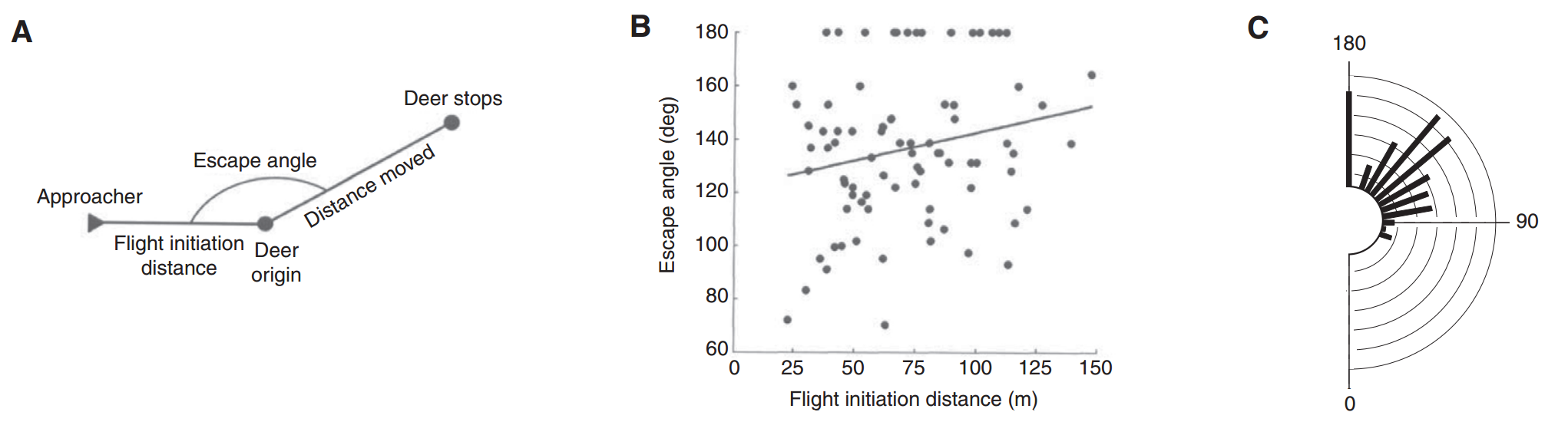}
\caption{\revisiontwo{‘{\it a)} The typical parameters to describe the flight of a black-tailed deer with {\it b)} measured escape angel as the function of  flight initiation distance and {\it c) } escape angle frequency distribution.’} \cite{domenici2011animal}.}
\label{fig:deer}
\end{center}
\end{figure} 

\revisiontwo{When the escaping is erratic, the prey can use a mixture of direct escaping and other, stochastic patterns of motion. This can be zigzagging which is a series of sudden changes in the direction or some even more advanced movements such as spinning, looping, and whirling (Figure \ref{fig:zigzag}) \cite{humphries1970protean, edut2004protean}.  For example, deer use a particular  distribution of escaping angles, which is also slightly dependent on the distance from its chaser (Figure \ref{fig:deer}).}
The effectiveness of these quick and unexpected changes in the escapers direction relies \revisiontwo{mostly} on the animals' inertia (and the difference between inertia of the prey and the predator) and the different kinds of time delays within the animals' sensory and motor system, even if their reaction time can be very short (in the order of magnitude of 50 $\mu$s) in chase and escape situations \cite{eaton2013neural}. 

\begin{table}
\centering
\label{table:table2}
\begin{tabular}{@{}ll@{}}
\textbf{Species} & \textbf{Escaping } \\
\textbf{} & \textbf{distance {[}m{]}} \\
\hline
Pronghorn & 235 \\
Mule deer & 149-250 \\
Elk & 85-201 \\
Bison & 101 \\ 
\end{tabular}
\caption{The escaping distance of several typical prey animals of large carnivores \cite{bentrup2008conservation}. }
\end{table}

In conclusion, we can say that the evaders are not always escaping directly from their chaser but are likely to combine direct escaping with some sort of random motion in order to trick their chasers and prevent their tactics (e.g. zigzagging) to be learnt. The direction of these random segments of their trajectory vary significantly among different species. The prey have typical escaping distances usually around a few hundred meters and they take into account all the chasers within that range when determining the direction and strategy of the escape (\revision{Table} \textcolor{blue}{1}) \cite{domenici2011animal2, bentrup2008conservation, cooper1997escape}.

\subsection{Previous models of prey-predator systems}

The latest paradigm of prey-predator modelling is the agent-based approach in which the main objective for chasers is to locate and catch the escapers as quickly as possible and for escapers to flee and avoid getting caught. The first agent-based theories were completely analytical, using the tools of different branches of mathematics from game theory to geometry but lacking many real-life biologically relevant attributes (e.g. noise) \cite{howland1974optimal, weihs1984optimal}.

Later on, the researchers have started studying the scenario of many chasers versus an escaper on a finite lattice with periodic boundary conditions, random spatial distribution of the agents, different sight ranges and uniform speed for all agents \cite{oshanin2009survival}. In the first models there was no interaction between the chasers and they were performing nearest-neighbour random walks to find the prey.

The rarely studied case of a faster escaper appeared in the article of Jin and Qu, in which they were studying how a group of pursuers can catch a faster evader in infinite open space when the chaser's finite sight range is bigger than the escaper's finite sight range \cite{jin2010pursuit}. They give geometric solutions on the question how the chasers can catch the prey by blocking it's escaping paths - a solution very sensitive to the initial spatial distribution. 

A recent agent-based group chasing model with many chasers and escapers was published by Kamimura et.al. \cite{kamimura2010group, vicsek2010statistical}. Here the group of chasers is pursuing the escapers on a two dimensional lattice with periodic boundary conditions. Because the chasers and the escapers have similar speed, the chasers can only catch an evader if they encircle it, which depends on the initial positions. They also defined a cost function to quantify the pursuit with which they determined the optimal number of chasers for a given number of escapers. 

Angelani's model has gone further than many previous ones because instead of a cell automata he constructed a model based on self-propelled agents according to the Vicsek-model \cite{angelani2012collective, vicsek1995novel}. The interactions between agents are formulated as physical forces in this two dimensional, continuous model with periodic boundary conditions, where the groups have similar speed. The agents' velocity vectors are calculated at each time step according to an alignment rule, a short range collision-avoidance force within the same groups, some noise and the chasing and escaping rules. This escaping tactics had only one parameter which also had an optimum with regards to the escapers.


\section{Model description}

\subsection{Simulation framework}

For studying group chasing we used a simulation framework developed in our research group, which was designed to model real-life agents and study their collective motion \cite{viragh2014flocking, vasarhelyi2014outdoor}. Because of their importance and generality in various prey-predator systems we added the following features of the framework to the group chasing model:

\begin{itemize}
\item {\bf Inertia } -- The restricted manoeuvrability and finite acceleration \revision{($a_{\rm max}$)} is a fundamental property of every moving object with finite size and mass. Therefore, it is crucial to include it in the model (note that the previously published chasing models were lacking this feature). In our framework it is assumed that the agents reach their desired velocity exponentially with a characteristics time $\tau_{\rm CTRL}$.

\item{\bf Time delay} -- In both biological and mechanical systems time delay is an unavoidable consequence of the limited speed of data transmission and processing (either via nerve cells and axons or electric wires and circuits). The simplest approach to describe it is to consider the value of a general time delay to be constant ($t_{\rm del}$).

\item{\bf Noise} -- The natural systems are usually biased by some kind of noise. To test our model's tolerance against such effects, a delta-correlated (Gaussian) outer noise term ${\bf \eta}(t)$ was added with standard deviation $\sigma$. 
\end{itemize}
\revision{Therefore, the general parameter-set describing a realistic agent (without the model specific parameters) is
\begin{eqnarray}
\{\tau_{\rm CTRL}, a_{\rm max}, t_{\rm del}, \sigma \}.
\end{eqnarray}
}

We are using velocity-based dynamic equations which are specified by the individual models. Such a velocity term\revision{, the desired velocity ${\bf v}^{\rm d}$ of}  the $i$th agent can be \revision{expressed in general with} the agent's position and velocity ${\bf r}_i$, ${\bf v}_i$ as the following:
\begin{eqnarray}
{\bf v}_i^{\rm d} (t) &=& {\bf f}_i(\{{\bf r}_i \}_{j=1}^N,\{ {\bf v}_j \}_{j=1}^N, {\bf \eta}(t)).
\end{eqnarray}
From this, the equation of motion is the following:
\begin{eqnarray}
{\bf a}_i &=& {\bf \eta}(t) + \frac{{\bf f}_i(...) - {\bf v}_i (t) }{|{\bf f}_i(...) - {\bf v}_i (t) |} \  \min \bigg\{ \frac{{\bf f}_i
(...) - {\bf v}_i (t) }{\tau_{\rm CTRL}}, a_{\rm max} \bigg\},  \label{eq:acc}\\ \nonumber \\
{\bf f}_i(...) &=& {\bf f}_i \big(\{ {\bf r}_j(t-t_{\rm del}) \}_{j 
\ne i}, {\bf r}_i(t), \{ {\bf v}_j(t-t_{\rm del}) \}_{j \ne i}, {\bf v}_i(t) \big). \label{eq:force}
\end{eqnarray}
\revision{Equation \ref{eq:acc} describes the acceleration characteristics of the $i$th agent reaching its desired velocity assuming finite acceleration, inertia and external noise. Equation \ref{eq:force} incorporates the assumption that the information from agent $j$ reaching agent $i$ is delayed \revision{by the given time parameter \revisiontwo{$t_{\rm del}$}}. }

\subsection{The realistic group chasing model}

\subsubsection{General properties of the model}

\paragraph{Agents}

In our model there are $N_{\rm c}$ chasers and $N_{e}$ escapers with different top speeds \revision{($v_{\rm max, c}$ $v_{\rm max, e}$)} but a similar top acceleration ($a_{\rm max}$). Let ${\bf r}_{{\rm c},i}, {\bf v}_{{\rm c},i}$ and ${\bf r}_{{\rm e},j}, {\bf v}_{{\rm e},j}$ be the $i$th chaser's and $j$th escaper's positions and velocities. An actively escaping prey will become inactive (out of the game) if the distance between it and the nearest chaser decreases below the catching distance ($r_{\rm cd}$); otherwise the agent stays active. In this model the agents are represented by a circular/spherical object with a radius of \revision{$r_{\rm cd}/2$}. The experiment is terminated when all of the escapers are caught.

\paragraph{Arena}

We aimed to build a bio-inspired model, therefore, we study group chasing within finite boundaries, because the pursuits in nature have boundaries both in space and time. However, many previously published models use either periodic boundary conditions or infinite space. Although both can be very useful e.g., in statistical physics, in the modelling of group chasing they can lead to very strange and unrealistic situations (for example, a prey ``coming back'' from the other side of the cell ``falling into the arms'' of the chasers wandering at the edge of the cell). For the sake of simplicity and to avoid edge-effects as much as possible, our simulation is running within a circular arena with the radius $r_{\rm a}$. We focus on the pursuit, therefore, we assume that each and every agent can see all the other agents -- we do not examine (group) foraging methods. In this constellation, the origin is the center of the arena.

The arena is surrounded by a soft wall \cite{han2006soft}. In this construction if any of the agents gets outside of the arena, a virtual agent starts to repulse it towards the center of the arena:
\begin{eqnarray} 
{\bf v}_i^{\rm a} &=& \revision{s} (r_i, r_{\rm a} , r_{\rm wall}) \  \bigg(v_{{\rm max}, k}
\  \frac{{\bf r}_i}{r_i} + {\bf v}_i \bigg) \qquad (k = {\rm e, c}),
\end{eqnarray} 
where 
\begin{eqnarray}
\hspace{-1.5cm}s(r_{i}, r_{\rm a}, r_{\rm wall}) = \cases{
0 & \textrm{if $r_{i} \in [0,r_{\rm a}]$}, \\ 
- \frac 12 \  \sin \bigg(\frac{\pi}{r_{\rm wall}} (r_{i}-r_{\rm a}) - \frac{\pi}{2} \bigg) - \revision{\frac12} & \textrm{if $r_{i} \in [r_{\rm a}, r_{\rm a} + r_{\rm wall}]$}, \\
- 1 & \textrm{if $r_{i} > r_{\rm a} + r_{\rm wall}$}
}
\end{eqnarray}
\revision{where} $r_{\rm wall}$ is the width of the wall and $r_i = |{\bf r}_i|$.

\paragraph{Collision avoidance}

Between each pair of same type agents there is a short-range repulsive interaction term to avoid collisions:
\begin{eqnarray} 
{\bf v}_{k,i}^{\rm coll} &=& v_{{\rm max},k} \  \mathcal{N} \Bigg[ \sum_j^{\revision{N_k}} \  \frac{|{\bf d}_{ij}| 
- \revision{{r_{\rm cd}}}}{|{\bf d}_{ij}|} \  {\bf d}_{ij} \  \Theta(\revision{{r_{\rm cd}}}- |{\bf d}_{ij}|) \Bigg], \quad {\rm where} \\
{\bf d}_{ij} &=& {\bf r}_{k, i} - {\bf r}_{k, j} \qquad (k = {\rm e, c}),
\label{eq:rep} 
\end{eqnarray}
\revision{$r_{\rm cd}$ is the catching distance defined as two times the radius of the agents, }  $\Theta$ is the Heaviside step function, ${\bf d}_{ij}$ \revision{is} the displacement vector between the $i$th and $j$th agent, and the $\mathcal{N}[.]$ operator normalizes its argument and returns it as a dimensionless vector.

\paragraph{Viscous friction-like interaction}

To simulate realistic motion, we introduced viscous friction-like interaction terms \cite{cucker2007emergent}, parameterized with a specific friction coefficient $C_{\rm f}$, with the following general form
\begin{eqnarray} 
{\bf v}_{ij} &=& C_{\revisiontwo{\rm f}} \  \frac{{\bf v}_j - {\bf v}_i }{({\rm max}\{\revision{{r_{\rm cd}}}, |{\bf d}_{ij}| \})^2}.  
\label{eq:frrr} 
\end{eqnarray}
Additionally, viscous friction\revision{-like} damping is important because it minimizes unwanted oscillations in any delayed system \cite{viragh2014flocking}.

\subsubsection{Chasers}

\paragraph{Chasing rule} Each chaser chases the closest escaper \revision{(chasing term)}. \revision{The collective chasing} strategy includes a soft repulsion between \revision{the chasers} and the prediction of their target's position.

\paragraph{Chasing  term} 

\begin{figure}
\begin{center}
\includegraphics[width=0.65\textwidth]{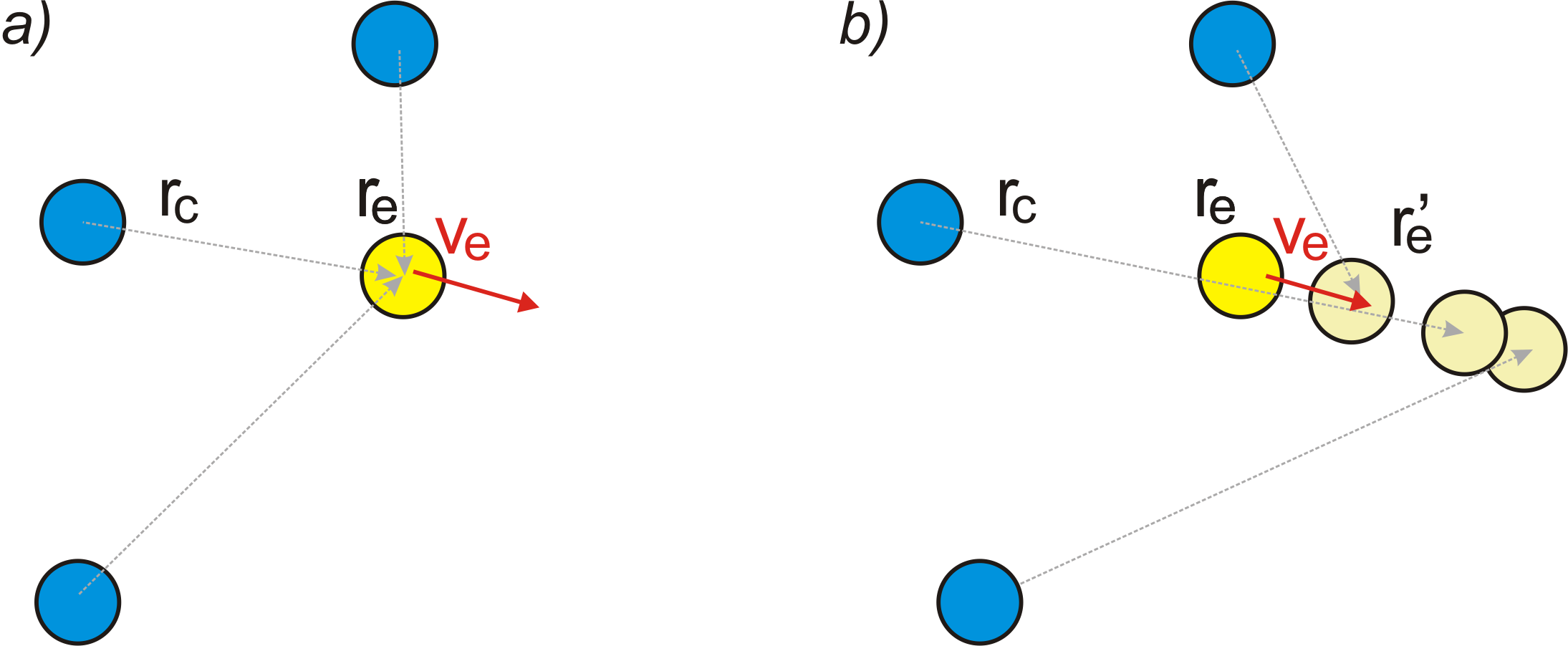}
\caption{The chasers (blue) are chasing the escaper's (yellow) {\it a)} current {\it b)} predicted position according to (\ref{eq:pred}).}
\label{fig:prediction}
\end{center}
\end{figure} 
The chasing is represented by an attractive interaction between the $i$th chaser and its target escaper toward the escaper \revision{and a  viscous friction-like velocity alignment term} in the following way:
\begin{eqnarray}
{\bf v}_{{\rm c}, i}^{\rm ch} &=& v_{\rm max,c} \  \mathcal{N}\Bigg[ \frac{{\bf r}_{{\rm c},i} - {\bf r}_{\rm e}}{|{\bf r}_{{\rm c},i} - {\bf r}_{\rm e}|} - C'_{\rm f} \  \frac{{\bf v}_{{\rm c}, i} - {\bf v}_{\rm e} }{|{\bf r}_{{\rm c},i} - {\bf r}_{\rm e}|^2} \bigg],
\label{eq:chasing}
\end{eqnarray}
\revision{where $C'_{\revisiontwo{\rm f}}$ is the coefficient of the  velocity alignment term relative to the attraction.}
\paragraph{Prediction} 
A part of the chasers strategy is the cognitive prediction of their prey's future position. This means that the chasers are heading to the point ${\bf r'}_{\rm e}$ (the estimated future coordinate of the escaper) instead of the escaper's current position ${\bf r}_{\rm e}$ (Figure \ref{fig:prediction}). Here we only use a simple linear approximation to determine the value of ${\bf r'}_{\rm e}$, which is the point at which the chaser could catch the escaper if the time needed for this is below a certain threshold time ($\tau_{\rm pred}$); if this is not the case, this is the point which the chaser can reach during $\tau_{\rm pred}$ time. Then the value of ${\bf r'}_{\rm e}$ is the numeric solution of the following equation:

\begin{eqnarray}
{\bf r'}_{\rm e} &=& {\bf r}_{\rm e} + \revision{   \frac{{\bf v}_{\rm e}}{|{\bf v}_{\rm e}|}} \  {v}_{\rm max, e} \  {\rm min} \bigg[\frac{{\bf r'}_{\rm e} - {\bf r}_{\rm c}}{\revisiontwo{{v}_{\rm max, e}}}\revision{,} \tau_{\rm pred} \bigg]
\label{eq:pred}
\end{eqnarray}

\paragraph{Interaction between chasers} Let us assume that a group of chasers is chasing the same escaper. If there is no interaction between the chasers they will eventually get too close to each other because their driving forces are similar. This results in a tail of chasers behind the escaper. This is neither a logical, nor an observable predatory behaviour in nature. What we can see in nature though, is that chasers do not get too close to each other while they are pursuing their prey, but the distance between them decreases as they get closer to (and finally catch) the escaper. We modelled this with a repulsive interaction term which has a characteristic length of $r_{\rm inter}$,  a magnitude factor \revision{of} $C_{\rm inter}$, and a magnitude of $C_{\rm inter} \  v_{\rm max, c}$. Consequently, $C_{\rm inter} = 0$ means that there is no interaction and $C_{\rm inter} = 1$ means that the repulsion between chasers is just as strong as their attraction towards the escaper. The mathematical definition of this interaction is:
\begin{eqnarray}
\hspace{-1.5cm}{\bf v}_{{\rm c},i}^{\rm inter} &=& C_{\rm inter} \revision{{v}_{\rm max, c}} \  \mathcal{N} \bigg[ \sum_j \bigg( {\bf d}_{ij} \  \frac{| {\bf d}_{ij} | - r_{\rm inter}}{| {\bf d}_{ij} |}- C''_{\revisiontwo{\rm f}} \  \frac{{\bf v}_{\revision{{\rm c}}, i} - \revision{{\bf v}_{\revision{{\rm c}}, j} }}{| {\bf d}_{ij} |^2} \bigg) \bigg], \label{eq:inter}\\%
{\bf d}_{ij} &=& {\bf r}_{\revision{{\rm c}}, i} - {\bf r}_{\revision{{\rm c}}, j},
\end{eqnarray} 
\revision{where $C''_{\revisiontwo{\rm f}}$ is the viscous friction-like term's coefficient.}

The driving force of the $i$th chaser is the sum of the previously introduced terms (interaction with the wall of the arena, collision avoiding short-term repulsion, direct chasing force and the long-term repulsion between chasers):
\begin{eqnarray}
{\bf f}_{{\rm c}, i} &=& {\bf v}_{i}^{\rm a} + {\bf v}_{{\rm c}, i}^{\rm coll} + {\bf v}_{{\rm c}, i}^{\rm ch} + {\bf v}_{{\rm c}, i}^{\rm inter}.
\end{eqnarray}

\subsubsection{Escapers}

\paragraph{Escaping rule} The escaper moves towards the furthest free direction away from all the chasers within its sensitivity range. 
At the wall, the escaper aligns its velocity to the wall. If the escaper can slip away between the nearest two chasers, then gets back to the field. We defined a panic parameter which depends on the distance between the escaper and its nearest chaser and controls the erratic behaviour. 

\paragraph{Direct escaping} An escaping agent takes into account every chaser within its range of sensitivity ($r_{\rm sens}$), while it weights this effect by the distance of each chaser. This represents the biological observation that the prey cares about all the chasers which are too close and the closer a predator is, the more dangerous it is. This interaction term for the $i$th escaper is the following:

\begin{eqnarray}
\hspace{-1.5cm} {\bf v}_{{\rm e}, i}^{\rm esc} &=& v_{\rm max, e} \  \mathcal{N} \bigg[ \sum_j \bigg( \frac{{\bf r}_{{\rm e},i} - {\bf r}_{{\rm c},j}}{|{\bf r}_{{\rm e},i} - {\bf r}_{{\rm c},j}|^2} - C'''_{\rm f} \frac{{\bf v}_{{\rm e},i} - {\bf v}_{{\rm c},j}}{|{\bf r}_{{\rm e},i} - {\bf r}_{{\rm c},j}|^2}\bigg) \Theta(r_{\rm sens} - |{\bf r}_i - {\bf r}_j|) \bigg],
\end{eqnarray}
\revision{where $C'''_{\revisiontwo{\rm f}}$ is the coefficient of the velocity alignment term.}

\paragraph{Erratic escaping} 
In nature many species tend to use certain kinds of erratic escaping strategies in which they combine direct escaping with some more advanced patterns of motion. Here we implemented the most basic one called zigzagging \cite{edut2004protean, humphries1970protean}. During zigzagging, the escaper tries to trick its chaser(s) with a set of sudden and unexpected changes in its escaping direction. For this we introduced a so-called panic parameter ($p_{\rm panic}$) which is an exponential function of the distance between the escaper and the nearest chaser ($d_{\rm min}$) with the value of 0 if the chaser is outside of the escapers range of sensitivity and a value of 1 when the escaper gets caught:
\begin{eqnarray}
p_{\rm panic} &=& \frac{1}{e - 1} \  \bigg[ e^{-d_{\rm min}/r_{\rm sens} +1} - 1 \bigg]. \label{eq:panic}
\end{eqnarray}

The escaper starts zigzagging when the panic parameter reaches a certain threshold ($p_{\rm thres}$) and it is not closer to the wall than a certain distance ($r_{\rm zigzag}$):
\begin{eqnarray}
p_{\rm panic} &<& p_{\rm thres}, \\ r_{\rm e} &<& r_{\rm a} - r_{\rm zigzag},
\end{eqnarray} 
where $r_{\rm e}$ is the absolute value of ${\bf r}_{\rm e}$. If these conditions are true, the escaper starts moving in a certain direction for a given amount of time. Afterwards, the prey zigzags in another direction for a period of time if the zigzag conditions are still true, otherwise it continues to escape directly. The zigzagging gets interrupted when the distance between the chaser and the escaper decreases below $r_{\rm zigzag}$. We determine the direction of a zigzag segment in the following way:

\vspace{-0.5cm}
\begin{eqnarray}
{\bf v}_{\revision{\rm e}, i}^{\rm zigzag} = v_{\rm max, e} \  \mathcal{N} \bigg\{ \mathcal{R} \bigg[ \sum_j \bigg( \frac{{\bf r}_{{\rm e},i} - {\bf r}_{{\rm c},j}}{|{\bf r}_{{\rm e},i} - {\bf r}_{{\rm c},j}|^2} \bigg) \  \Theta(r_{\rm sens} - |{\bf r}_i - {\bf r}_j|) \bigg] \bigg\},
\end{eqnarray}
where $\mathcal{R} $ is a rotating operator which rotates its argument vector by a randomly chosen angle in two dimensions from the $[-\pi, \pi]$ interval and by two random (an Azimuth and a polar) angles in three dimensions from the intervals of $[0, 2\pi]$ and $[0, \pi]$. The duration of a zigzag segment is also a random parameter which we chose from a power-law distribution with an exponent of $-2$ with a lower limit of $r_{\rm zigzgag}/v_{\rm max,e}$ and an upper limit of $r_{\rm a}/v_{\rm max,e}$.

\paragraph{Behaviour at the wall} Confronting the wall would significantly decrease the escaper's speed; therefore, the escaper should avoid getting too close to the wall. This is ensured by cutting off the escaper's velocity vector's radial component depending on how close to the wall the agent is:
\begin{eqnarray}
{\bf v}_{\rm e}^{\rm final} & = & {\bf v}_{\rm e} - C_{\rm wall}(r_{\rm e}, r_{\rm a}, r_{\rm wall}) \  \frac{{\bf r}_{\rm e}}{{r}_{\rm e}}   \bigg( \frac{\revisiontwo{{\bf r}_{\rm e}^{\rm T}}}{{r}_{\rm e}}  \revision{\cdot}  {\bf v}_{\rm e} \bigg) = \hat{\rm \bf W} {\bf v}_{\rm e}, \label{eq:walll}
\end{eqnarray}
where $C_{\rm wall}$ is defined as the following:

\begin{eqnarray}
\hspace{-2.0cm}C_{\rm wall} = \cases{
1 & \rm{if $r_{\rm e} > r_{\rm a} - 2 r_{\rm wall}$}, \\ 
\frac12 \bigg[ 1 - \sin \bigg(\pi \  \frac{r_{\rm e} - r_{\rm a} + 2r_{\rm wall}}{r_{\rm zigzag}} - \frac{\pi}{2} \bigg) \bigg] & \rm{if $r_{\rm e} > r_{\rm a} - 2r_{\rm wall} - r_{\rm zigzag}$},\\
0 & \rm{otherwise.}
}
\end{eqnarray}

\begin{figure}[hbt]
\begin{center}
\includegraphics[width=1.0\textwidth]{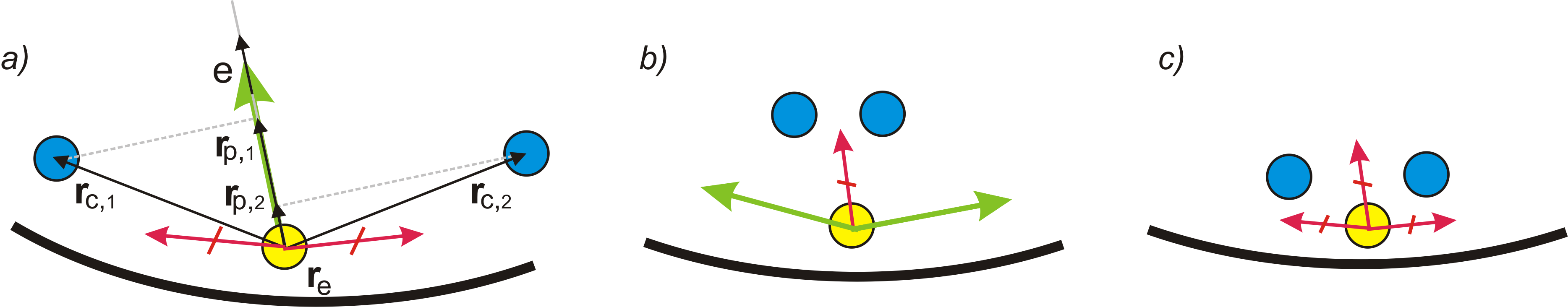}
\caption{ {\it a)} The escaper (yellow)  can slip through the gap (according to (\ref{eq:tauu})) between the chasers (blue), therefore it escapes (green arrow), but if it stayed at the wall, would get caught (red arrows) {\it b)} the escaper cannot slip through the gap between the chasers but can escape by aligning to the wall  {\it c)} the chasers are blocking all the possible escape directions.   }
\label{fig:wall}
\end{center}
\end{figure}

Applying exclusively this method would mean that once the escaper reaches the wall gets stuck there forever, which would be unrealistic. Thus, we extended this rule: if the escaper is close to the wall ($C_{\rm wall} > 0.5$) and it can slip through the gap between the two closest chasers it does, and gets back to the center part of the arena. In order to determine whether or not jumping back to the arena means successful escaping, the prey routinely calculates how long it would take for itself and the two nearest chasers to reach certain points. These points are the ones we get when we project the difference vectors pointing from the prey to the chasers onto the line defined by the sum of these two vectors ((\ref{eq:eee}), Figure \ref{fig:wall}).

\begin{eqnarray}
{\bf e} &=& \frac{ ({\bf r}_{{\rm c},1} - {\bf r}_{\rm e}) + ({\bf r}_{{\rm c},2} - {\bf r}_{\rm e} ) }{ |({\bf r}_{{\rm c},1} - {\bf r}_{\rm e}) + ({\bf r}_{{\rm c},2} - {\bf r}_{\rm e} )| }, \label{eq:eee} \\
{\bf r}_{{\rm p},k} &=& {\bf e}   \big[ \revision{\bf e^T}  \revision{\cdot}  ({\bf r}_{{\rm c},k} - {\bf r}_{\rm e}) \big], \\
\tau_{{\rm e}, k} &=& \frac{r_{{\rm p},k}}{v_{\rm max, e}}, \\
\tau_{{\rm c}, k}&= &\frac{\revisiontwo{|  {\bf r}_{{\rm p},k} -{\bf r}_{{\rm c},\revisiontwo{k}}+ {\bf r}_{\rm e}|}  - \revision{ r_{\rm cd}}}{v_{\rm max, \revisiontwo{c}}}, \\
\tau_{{\rm e}, k} &\stackrel{?}{<}& \tau_{{\rm \revisiontwo{c}}, k}, \label{eq:tauu}
\end{eqnarray}
where $k = 1, 2$ are the indices of the two nearest chasers (see on Figure \ref{fig:wall}). If the (\ref{eq:tauu}) condition is true, the escaper can run back to the arena without getting caught; otherwise it stays at the wall where it will either be caught or chased.

The driving force of the $i$th escaper will be the sum of the interaction with the wall, collision avoiding repulsion and either the direct or the zigzagging escaping term, at the end transformed by the $\hat{\rm \bf W}$ operator (\ref{eq:walll}):
\begin{eqnarray}
{\bf f}_{{\rm e}, i} &=& \hat{\rm \bf W} \big({\bf v}_i^{\rm a} + {\bf v}_{{\rm e}, i}^{\rm coll}+ ( {\bf v}_{{\rm e}, i}^{\rm esc} \lor {\bf v}_{{\rm e}, i}^{\rm zigzag} ) \big).
\end{eqnarray}

\subsubsection{Model parameters}

Although in our model the number of expressions and parameters is significantly larger than in the simplest statistical mechanical models, these expressions and parameters are far from being arbitrary, almost all of them are in a direct relation with the observed systems and are originating from such trivial facts as the existence of time delays (e.g. reaction time). The parameters are "freely" tuneable from the point of the model, however, most of them are experimentally measurable for an actual predator-prey system.

 These parameters can  be separated into two sets: environmental and tactical parameters (\revision{Table} \textcolor{blue}{2}). Except $t_{\rm del}$, all the environmental parameters had fixed values during the measurements, for instance, $a_{\rm max} = 6$ m/s$^2$, $r_{\rm a} =$ 150 m and $ r_{\rm wall} = 5$ m. On the other hand, all the tactical parameters except the fixed velocities ($v_{\rm max,c}= 6$ m/s, \revision{$v_{\rm max,e} =  8$ m/s}) can be adjusted at the beginning of each run to create and study different chasing and escaping tactics. As it will be remarked in the text, we studied the most of the cases both in 2D and 3D.

\begin{table}[]
\label{table:params}
\begin{tabular}{lccl}
\textbf{\begin{tabular}[l]{@{}l@{}}Environmental\\ parameters\end{tabular} }  & \textbf{\revision{Typical range}} & \textbf{Dimension}  &  \textbf{Definition} \\
$a_{\rm max}$  & \revision{  6  }      &            m/s$^2$&       Maximum acceleration of the agent \\
$\sigma$ &  \revision{  0.0-1.0  }   &           m$^2$/s$^3$ &    Standard deviation of the Gaussian noise 
\\
$\tau_{\rm CTRL}$&  \revision{ 0.06   }          & s & Characteristic time of the acceleration \\
$t_{\rm del}$ &   \revision{  0-6  }         & s & Delay time \\
$C'_{\rm f}$&    \revision{  1.1  }         & ms & Friction coefficient in the direct chasing term \\
$C''_{\rm f}$&   \revision{  1.1  }          & m$^2$s & \revision{Friction coefficient in the velocity} \\ 
                     &                                             &               &    \revision{alignment between chasers} \\
$C'''_{\rm f}$ &   \revision{ 1.1   }         & s & Friction coefficient in the direct escaping term \\
$r_{\rm a}$ &   \revision{ 150   }        & m & Radius of the arena \\
$r_{\rm cd}$&   \revision{  1  }          & m & Catching distance \\
$ r_{\rm wall}$ &   \revision{  5  }         & m & Width of the wall \\
\end{tabular}

\vspace{8pt}
\label{table:params2}
\begin{tabular}{lccl}
\textbf{\begin{tabular}[l]{@{}l@{}}Chasers' \\tactics\\ parameters\end{tabular} }  & \textbf{\revision{Typical range}}  & \textbf{Dimension} & \textbf{Definition} \\
$v_{\rm max,c}$   &      \revision{    6}           & m/s & Maximum velocity of the chasers \\
$\tau_{\rm pred}$ &    \revision{  0-6   }     & s & Upper limit of the chaser's prediction \\
$C_{\rm inter}$ &    \revision{   0-1      }    & 1 & Interaction's strength between chasers \\
$r_{\rm inter}$&    \revision{    0-300    }    & m & Interaction distance between chasers \\
\end{tabular}

\vspace{8pt}
\label{table:params3}
\begin{tabular}{lccl}
\textbf{\begin{tabular}[l]{@{}l@{}}Escapers' \\tactics\\ parameters\end{tabular} } & \textbf{\revision{Typical range}}  & \textbf{Dimension} & \textbf{Definition} \\
$ v_{\rm max,e}$  &   \revision{   8        }    & m/s & Maximum velocity of the escapers \\
$p_{\rm thresh}$ &   \revision{     0-1    }      & 1 & Escaper's panic threshold \\
$ r_{\rm sens}$     &    \revision{    0-120    }      &m & Escaper's range of sensitivity \\
$ r_{\rm zigzag}$ &     \revision{  0-80     }       & m & \begin{tabular}[c]{@{}l@{}}Minimal length of a zigzag segment; the distance \\limit between an escaper and a chaser when the \\ escaper stops zigzagging\end{tabular}
\end{tabular}
\caption{The environmental and tactical parameters of the model.}
\end{table}


\section{Results and discussions}

We use finite and fixed simulation length ($\tau_{\rm max} = 600$ s) which is comparable to the observed length of pursuits in nature, and also practical for the simulations because of the limited computation time. This means that when counting the average results of the runs we have to handle properly those cases in which the escaper was not caught. Therefore we assume that the uncaught prey's lifetime was the maximum, so if the escaper's average lifetime is $\tau_{\rm esc}$ out of $n$ runs when it was caught, the average time is:
\begin{eqnarray}
\overline{\tau}_{\rm esc}&=& \frac{n \  \tau_{\rm esc} + (n_{\rm tot} - n) \  \tau_{\rm max} }{n_{\rm tot}}. \label{eq:efff}
\end{eqnarray}

We defined the following effectiveness functions to quantify the pursuit with regards to the chasers:
\begin{eqnarray}
{\rm Effectiveness_c} &= & \frac{n/n_{\rm tot}}{ \overline{\tau}_{\rm esc} / \tau_{\rm max} \  N_{\rm c}}, \label{eq:eff}
\end{eqnarray}
which means that the group of chasers is as effective as many times it catches the escaper out of $n_{\rm tot}$ same runs, gets the prey as fast as possible, and for this they need the smallest number of chasers. The division by $N_{\rm c}$ is referring to the observation that the more chasers catch the same prey, the less food per capita they get. On the other hand we defined an effectiveness function for escapers, as well:
\begin{eqnarray}
{\rm Effectiveness_e} & = & \frac{\overline{\tau}_{\rm esc}}{ \tau_{\rm max}}.
\end{eqnarray}
The escaper's effectiveness is the ratio of the average lifetime in maximum time allowed and those who are not caught are taken into account with Effectiveness$_{\rm e}$ of 1.

\subsection{Group of chasers versus a single escaper}

\subsubsection{Strength of the interaction and the number of chasers}

\begin{figure}
\begin{center}
\includegraphics[width=0.35\textwidth]{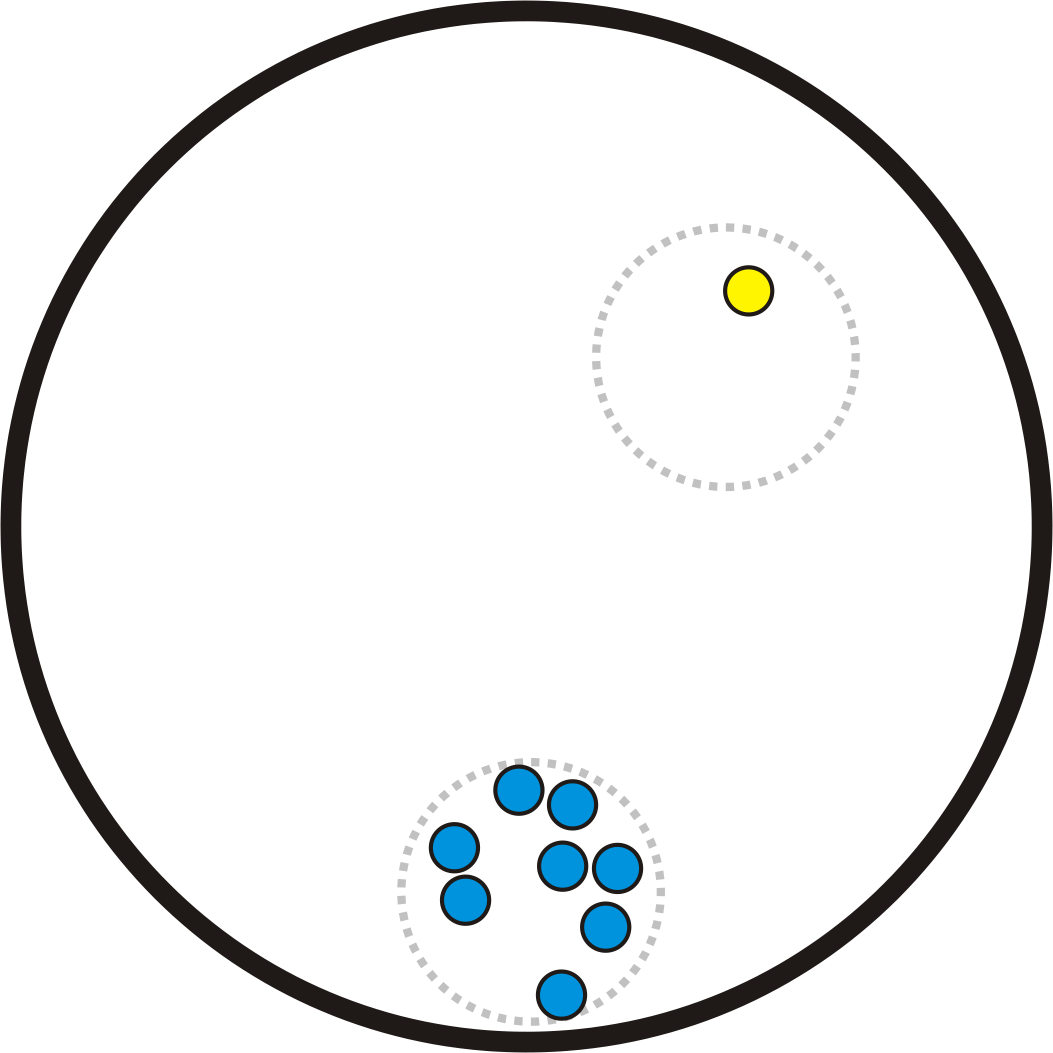}
\caption{Pack starting distribution of the agents - while the origin is the center of the field, the escaper (yellow) was placed randomly in a circle  (dashed line) with a radius of $r_{\rm a}/4$ and a center with the coordinates of ($r_{\rm a}/2$, $r_{\rm a}/2$)  while the chasers (blue) were put in another circle with the same radius but with a center of ($0$, $-3r_{\rm a}/4$).}
\label{fig:pack}
\end{center}
\end{figure}
At first we studied the effect of the interaction between the chasers as a function of $N_{\rm c}$ with two different spatial distributions of the agents and the interaction range fixed ($r_{\rm inter}$ = 300 m). In the first one, all the agents were spaced uniformly randomly on the field with no regards to their type. In the second one, the escaper was placed randomly in a circle with a radius of $r_{\rm a}/4$ and a center with the coordinates of ($r_{\rm a}/2$, $r_{\rm a}/2$) while the chasers were put in another circle with the same radius but with a center of ($0$, $-3r_{\rm a}/4$), where the center of the field is the origin (\revision{Figure \ref{fig:pack}}). In 3D the method is the same using spheres. The completely random starting position is widely used in the literature, however in reality it is more common that the chasers are closer to each other when they start the pursuit (it is also not logical from the prey to wander into a place where the predators are scattered all around).

\begin{figure}
\begin{center}
\vspace{-0.5cm}
\includegraphics[width=0.6\textwidth]{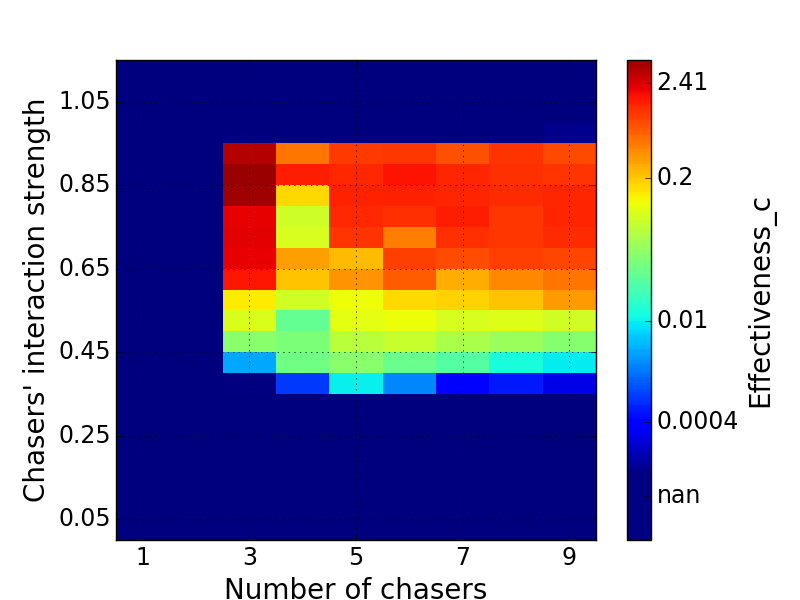}
\vspace{-0.05cm}
\caption{\revision{The effectiveness of the group of chasers in the case of pack starting distribution in 2D as the function of the number of chasers and the interaction's strength between them. In the dark blue regime the chasers never catch the prey. }} 
\label{fig:Fig2p}
\end{center}
\begin{minipage}[c]{0.49\textwidth}
\includegraphics[width=\textwidth]{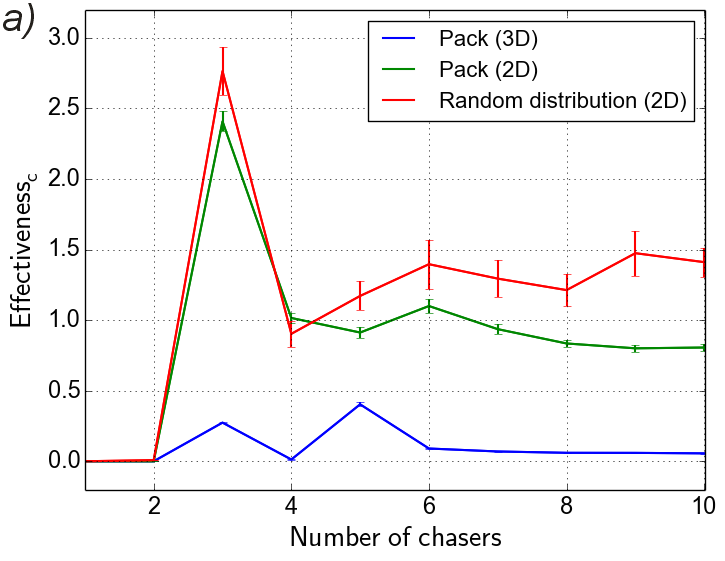}
\end{minipage}
\begin{minipage}[c]{0.5\textwidth}
\includegraphics[width=\textwidth]{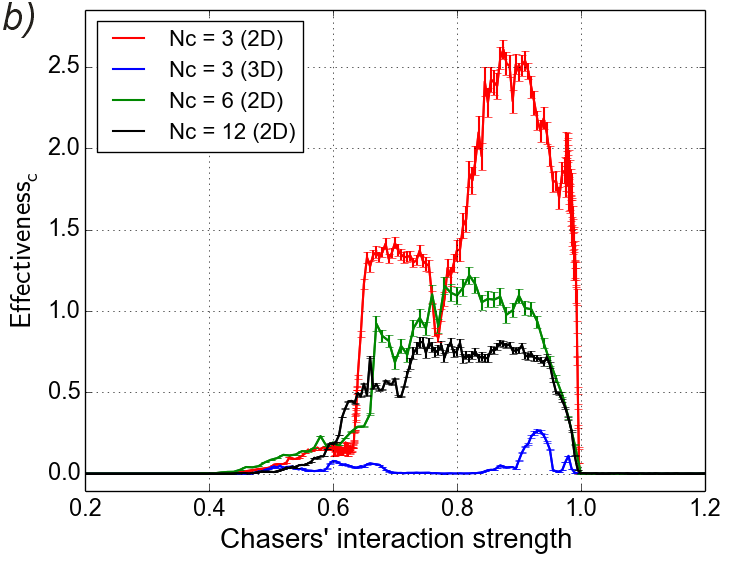}
\end{minipage}\hfill
\begin{minipage}[c]{0.49\textwidth}
\includegraphics[width=\textwidth]{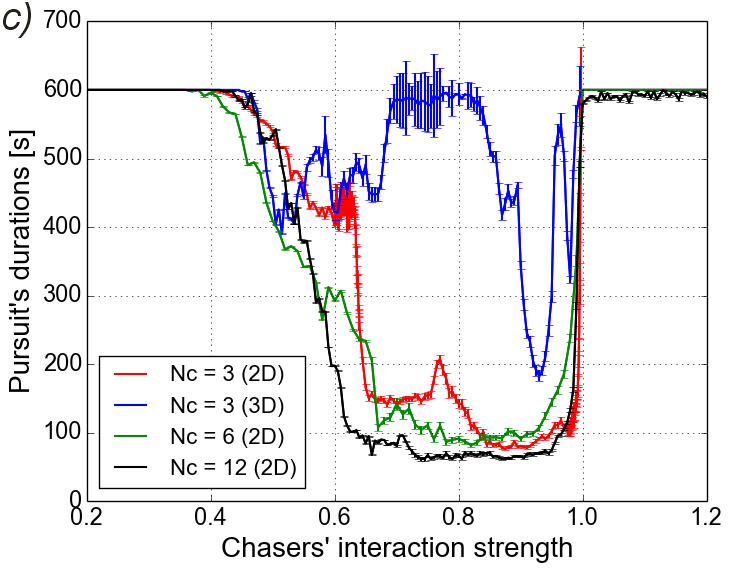}
\end{minipage}
\begin{minipage}[c]{0.5\textwidth}
\includegraphics[width=\textwidth]{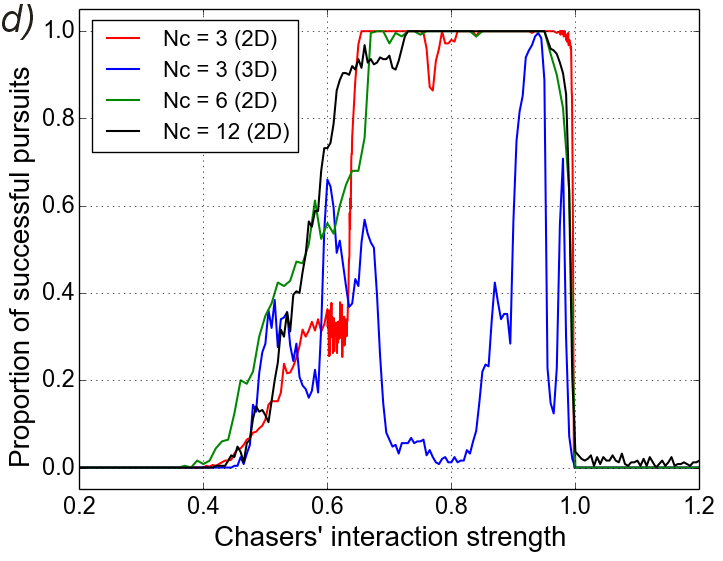}   
\end{minipage}\hfill
\vspace{-0.05cm}
\caption{\revision{Cross sections of Figure \textcolor{blue}{8}: the Effectiveness$_{\rm c}$ as a function of {\it a)} $N_{\revisiontwo{\rm c}}$ when $C_{\rm inter} = 0.9$ {\it b)}  $C_{\rm inter}$ when $r_{\rm inter} = 300$ m for various cases {\it c)} the pursuit's duration and  {\it d)} \revisiontwo{the} proportion of successful pursuits (which ended in catching the escaper). We can  see on the graphs that the cooperative hunting is effective with well-tuned interaction between the chasers,  and optimal when $N_{\revisiontwo{\rm c}} = 3$ for 2D, $N_{\revisiontwo{\rm c}} = 5$ for 3D. There are two well-defined regimes in {\it b)} in which the chasers are, and are not successful. The narrow boundaries between the regimes are the results of the fixed geometric  assumptions.  The quick breakdown at $C_{\rm inter}$ = 1 means that at that point the chasers align into a regular grid formation, which can not be changed by a single (driver) escaper.}}
\end{figure}

For random distribution, the chasers sometimes happened to be initially around the escaper which means an immediate encircling and a very quick (and unrealistic) pursuit. However, with pack distribution the chasers have to build up their chasing formation every time to catch the prey, which causes a well-defined regime of the parameter space in which the chasers are successful. \revision{Despite this}, there is no big difference in the maximal Effectiveness$_{\rm c}$ values in these two cases\revision{, as Figure \textcolor{blue} 8 demonstrates}. For the case of pack distribution, the chasers' effectiveness can be seen on Figure \ref{fig:Fig2p} \revision{as the function of the number of chasers and the interaction strength between them. It is important to point out that these results are the solid proof of the fact that in these scenarios (e.g. the escaper is faster) a single chaser is not able to catch the prey  and multiple agents only have the chance to catch the prey if there is a strong tactical interaction between them. As an example, we took a closer look on certain cross sections around the optimum of Figure \ref{fig:Fig2p}, which are presented on Figure \textcolor{blue}{9}.  Figure \textcolor{blue}{9} illustrates that chasing in three dimensions is a lot harder then in two dimensions and while for the former case three chasers can form an optimal group, in three dimensions five is more advantageous. }

\revision{On one hand, we can see that the effectiveness decreases monotonously above $N_{\revisiontwo{\rm c}} = 6$ which happens due to the division by $N_{\revisiontwo{\rm c}}$ in (\ref{eq:efff}); on the other hand, one and two chasers have no chance ever to catch the faster prey. Between these two regimes the tendency is not monotonous. This happens because of geometric reasons (symmetric arrangements), which will be averaged out for larger $N_{\revisiontwo{\rm c}}$ values: odd number of chasers have much bigger chance to block all the escaping directions, a gap remains much easier when the number of chasers is even (see Figure \ref{fig:wall}).}

\subsubsection{Prediction \& delay}

\begin{figure}[b]
\begin{center}
\includegraphics[width=0.6\textwidth]{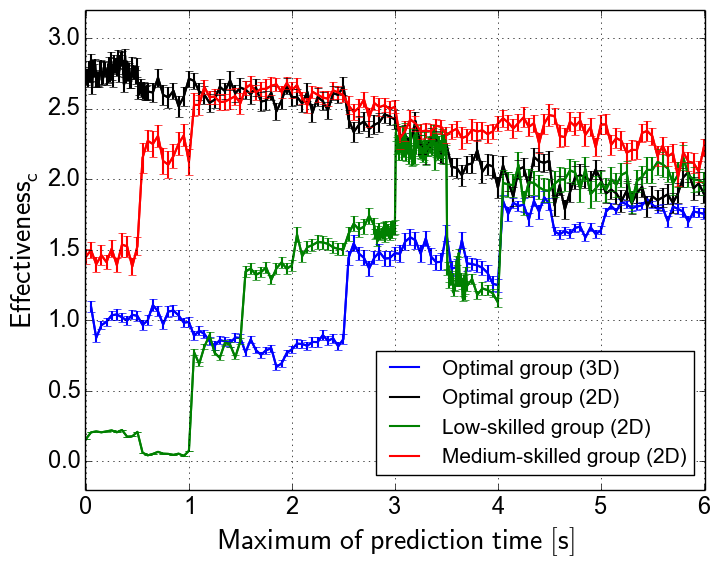}
\caption{The effectiveness of a previously optimized, a medium-level efficient (about half as effective as the optimized), and a low-level efficient (less effective than the 10\% of the optimized) group of three chasers as the function of the maximum of the prediction time. These results show that if the prediction time is large enough, it significantly enhances the ineffective group's effectiveness. On the other hand, in the case of the optimized group, longer prediction time expands the pursuit's duration, which (because of the 100\% success rate via the optimization), decreases the groups effectiveness.}
\label{fig:Fig9pred}
\end{center}
\end{figure} 

In this subsection we study the chaser's cognitive prediction regarding the escaper's path, how the time delay can affect the outcome of the pursuit, and how the prediction can compensate for the delay. As it was introduced in the model section, the prediction has an upper time limit, the maximum time of prediction ($\tau_{\rm pred}$) which is studied in this subsection. This quantity is proportional to how long the chasers forecast their prey's position. If we add the prediction (\revision{Figure} \textcolor{blue}{10}) to the previously optimized chasers, their effectiveness increases a little bit (with a maximum at $\tau_{\rm pred} \approx$ 0.38 s), but after that it starts to decrease. It happens because the optimized chasers are already hunting with a 100\% success rate ($n=n_{\rm tot}$ in (\ref{eq:eff})), while long prediction time extends the pursuit's duration because the chasers do not rush immediately towards to escaper but extrapolate and chase the escaper's future position which is always further and this longer duration causes the lower efficiency, while in 3D even an optimized (in regards to the interaction between the chasers) group's efficiency can be largely increased by the prediction. However, if we take a look on the chasing packs with non-optimized interaction parameters (in 2D), we can see that prediction improves their efficiently near to the level of the optimized one.

We tested the tolerance of the Effectiveness$_{\rm c}$ against delay with different prediction times for different groups of chasers (\revision{Figure} \textcolor{blue}{11}). These experiments led to the conclusion that if the prediction time is large enough in comparison to the delay (but still the same order of magnitude), prediction will compensate for the delay and the pursuit remains successful. However, very large delay (e.g. 5 s) with small prediction can completely prevent the success of the pursuit. Based on these results, the optimal values of the chasers' tactical parameters can be found in \revision{Table} \textcolor{blue}{3} if the system has a delay of 1 s.

\begin{figure}[h]
\begin{minipage}[c]{0.57\textwidth}
\hspace{-1cm}
\includegraphics[width=\textwidth]{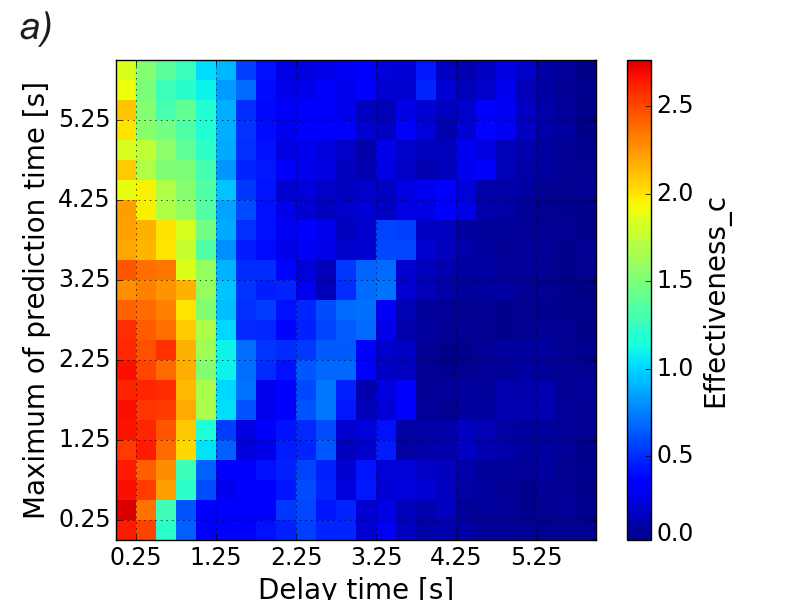}
\end{minipage}
\begin{minipage}[c]{0.57\textwidth}
\hspace{-1cm}
\includegraphics[width=\textwidth]{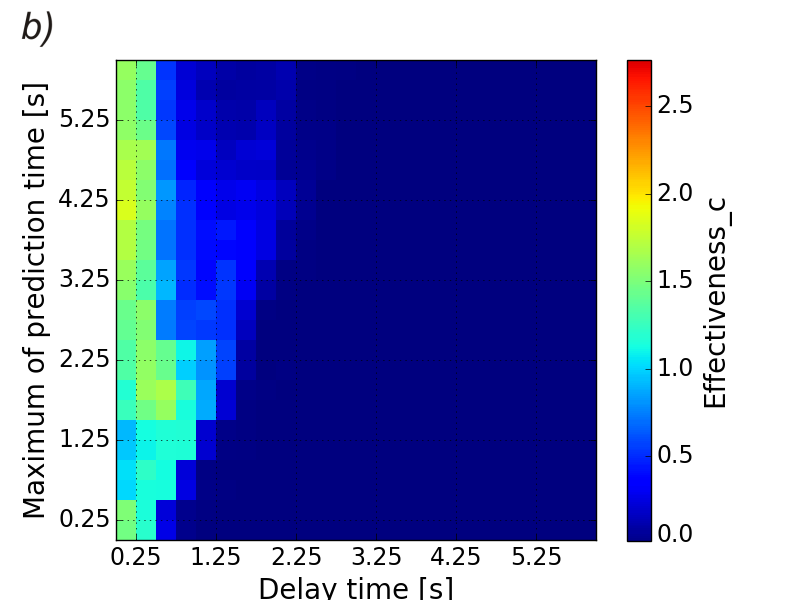}
\end{minipage}
\label{fig:Fig11}
\caption{\revision{Effectiveness of chasers as a function of maximum prediction time and delay time for an optimized group of three chasers in {\it a)} two and {\it b)} three dimensions.}}
\end{figure}

\begin{table}
\centering
\label{table:oop2}
\begin{tabular}{@{}l|cccc@{}}
& $N_{\rm c}$ & $C_{\rm inter}$ & $\revision{r}_{\rm inter}$ & $\tau_{\rm pred}$ \\ 
\hline
2D & 3 & 0.86 & 155 m & 1.75 s \\
3D & 3 & 0.9 & 135 m & 2 s \\ 
\end{tabular}
\caption{The optimal chasing parameters in two and three dimensions with prediction and delay (1 s).}
\end{table}

\subsubsection{The effect of outer noise}

We examined the effect of outer, Gaussian noise within the range of $\sigma = 0 - 2$ m$^2$/s$^3$, where 0.2 m$^2$/s$^3$ is equivalent to e.g. the blowing wind with a medium strength. The values of Effectiveness$_{\rm c}$ proved to be quite robust and  stable against these perturbation, just slightly increased the deviation of the statistical results. The standard deviation of the noise amplitude  --  Effectiveness$_{\rm c}$ dataset was about 2\% in two dimensions, and 2.4\% in three dimensions, is it stayed slightly constant.

\subsubsection{Emergent phenomenon in the model}

\begin{figure}
\centering
\includegraphics[width=\textwidth]{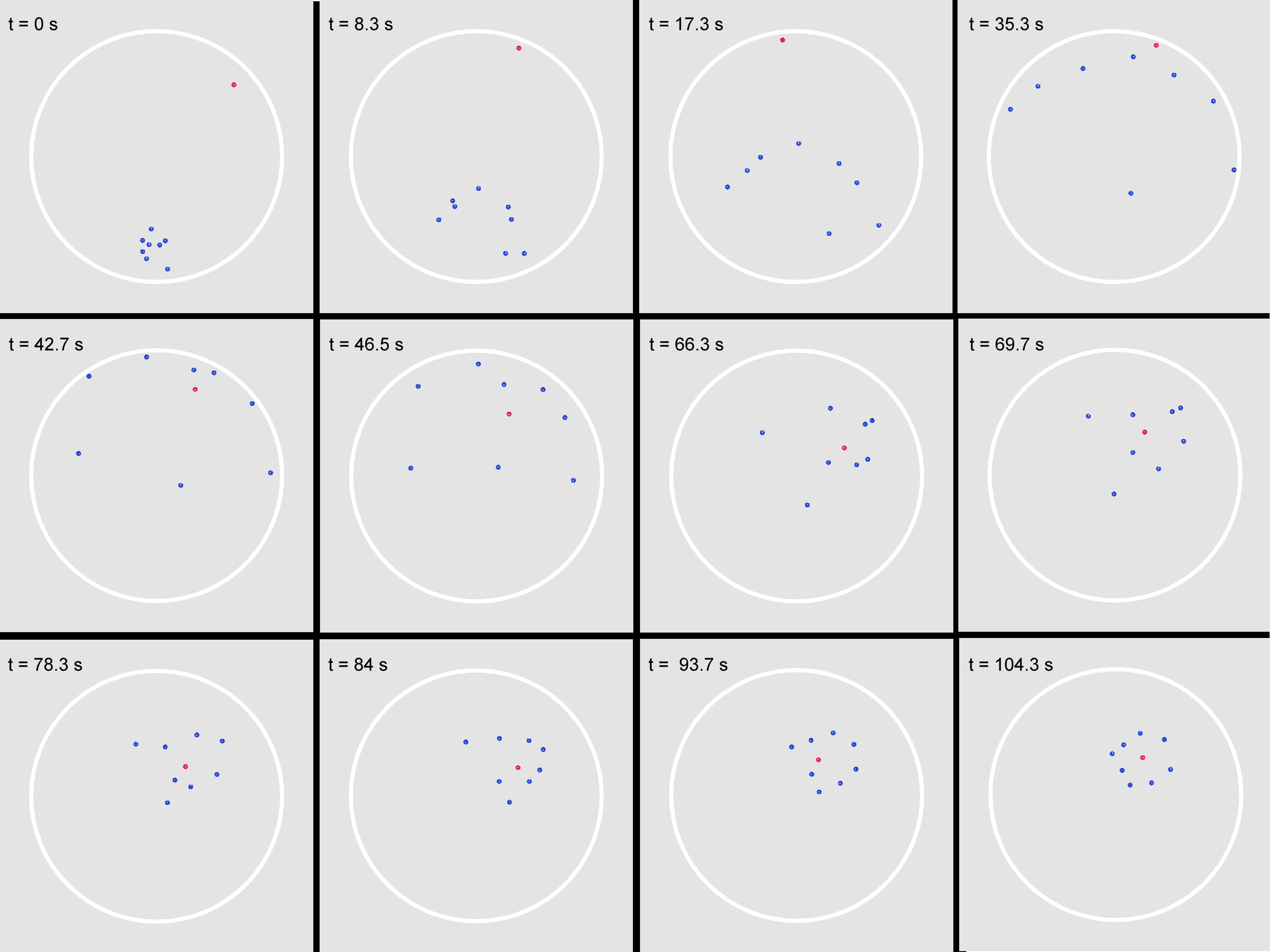} 
\caption{Nine chasers can quickly encircle the evader.}
\label{fig:sim}
\end{figure}

Despite our model's simplicity, while studying it in real time we can observe very interesting, life-like motion patterns emerging. These can be categorized as the following:
\begin{itemize}
\item {The optimal group of chasers catches the prey very quickly: they rush towards the escaper on the right trajectories to block all the possible escaping paths of the evader. This can also happen (with a lower frequency) in three dimensions. This sometimes results in line formation, too\revision{.}}
\item {If the chasing parameters are medium-level efficient,  ``on-the-field'' pursuit often happens: the agent go across the field several times, the escaper moves along the wall, and hops back to the center of the field several times. In these situations the classical encircling (observed e.g. about lions) can also happen  completely emergently (Figure \ref{fig:sim}). With significantly more chasers, this phenomenon (sometimes called as caging) can be observed in three dimensions as well.}
\end{itemize}

\subsubsection{Summary - chasers}

In this section we demonstrated that those chasers which solitarily are unable to catch the prey can successfully pursue it if there is a certain, well-parameterized interaction between them, which is their collective chasing tactics. For this we studied and optimized the group of chasers' parameters by mapping their parameter space. We have found that even in three dimensions three chasers can efficiently catch the prey if the interaction's strength and characteristic distance are set properly. We have shown that the prediction method enhances the capabilities of the chasers (even if their interaction parameters are sub-optimal), and that prediction can suppress the negative effects of time delay. We pointed out that these optimal chasing tactics are robust against external noise and have found interesting, emergent behaviour patterns, such as the life-like encircling. The data are also implying that chasing in three dimensions is a lot more difficult task in comparison to two dimensions, which is consistent with the fact that truly three dimensional pursuits are very rare in nature. \revision{What is more common in three dimensions is to restrict the degrees of freedom in some way. Flying and swimming animals frequently use the natural borders of their open 3D space (water-air or air-ground boundaries) to drive their prey into the corner. It has also been reported for group hunting dolphins that certain individuals tend to have different roles during the hunt, for example they are creating artificial blockades to restrict the possible escaping directions of the fish schools and catch them more efficiently \cite{norris1988cooperative,gazda2005division}. Finally, certain species, like the fish-hunting cone snails or all kinds of web spiders use various 'sit-and-wait' strategies -- they hide and attack abruptly only when the prey is close enough \cite{terlau1996strategy, kim2005cooperative}.}

\subsection{Erratic escaper}

We studied the panic threshold ($p_{\rm thresh}$) which is the control parameter of the zigzag pattern of motion and seek the optimum of $r_{\rm zigzag}$ and $r_{\rm sens}$. In this section $n_{\rm tot} = 1000$ for each data point.

\subsubsection{Zigzag motion: the panic threshold }

\begin{figure}[t]
\begin{minipage}[c]{0.5\textwidth}
\includegraphics[width=\textwidth]{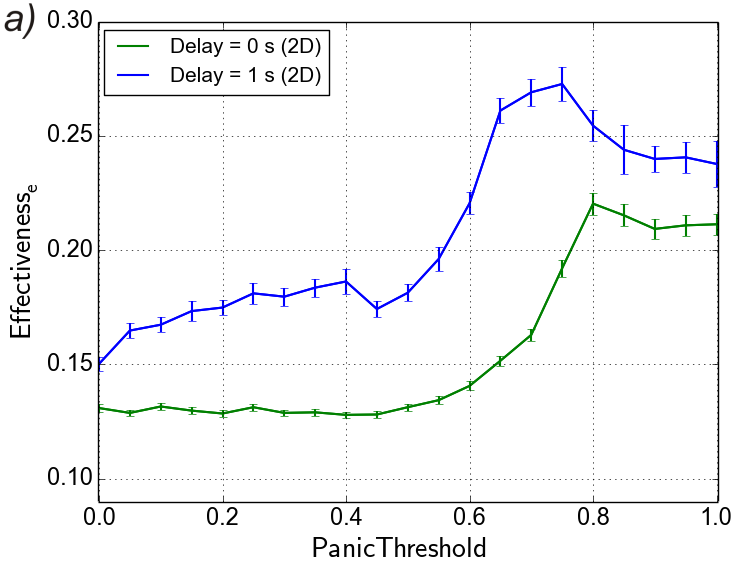}
\end{minipage}
\begin{minipage}[c]{0.5\textwidth}
\includegraphics[width=\textwidth]{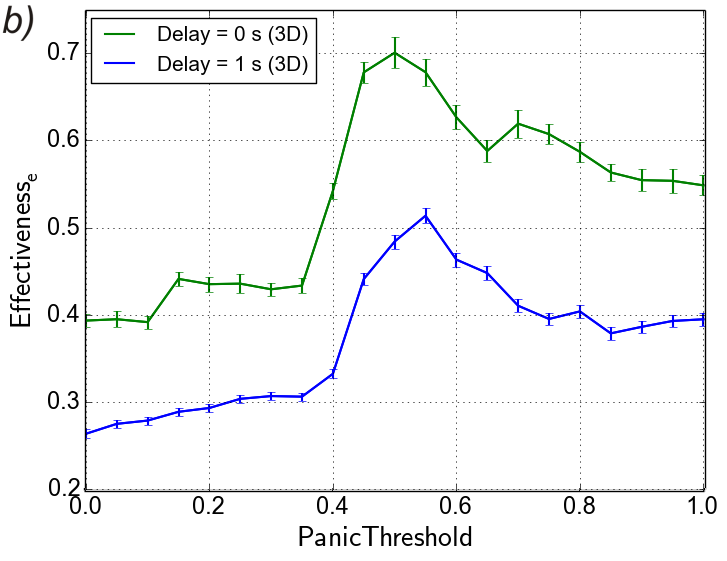}
\end{minipage}\hfill
\caption{The effectiveness$_{\rm e}$ as the function of $p_{\rm thresh}$ in {\it a)} two and {\it b)} three dimensions with and without delay shows that the escaper can optimize it's tactics with a well-chosen panic threshold parameter, therefore increase its lifespan.}
\label{fig:panicth}
\end{figure}

As the panic parameter of the zigzag motion was defined in (\ref{eq:panic}), it is a real number between 0 and 1 which depends on the distance between the escaper and its nearest chaser. If this value reaches the panic threshold, the escaper turns on the zigzag motion. At first we examined the Effectiveness$_{\rm e}$ values as a function of $p_{\rm thresh}$ to study the zigzag's effect on the pursuits outcome. The results are presented on Figure \ref{fig:panicth} with and without delay, in two and three dimensions. These figures tell us that zigzag can significantly improve the escaper's efficiency in all the studied cases, moreover, $p_{\rm thresh}$ does have optimal values.

\subsubsection{Zigzag motion: not too close }

\begin{figure}[h]
\begin{minipage}[c]{0.485\textwidth}
\vspace{-5pt}
\includegraphics[width=\textwidth]{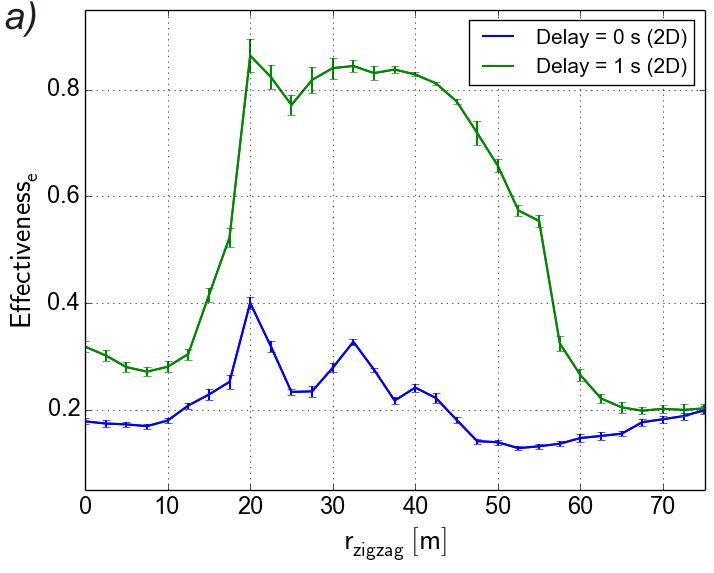}
\end{minipage}
\begin{minipage}[c]{0.5\textwidth}
\includegraphics[width=\textwidth]{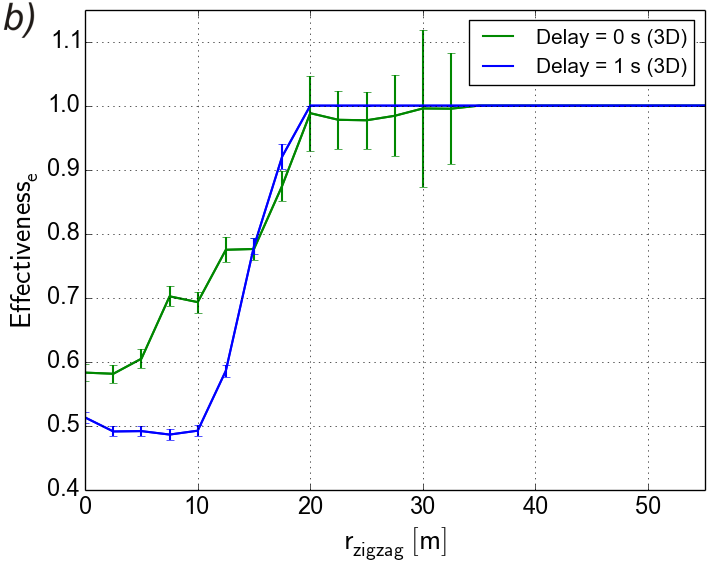}
\end{minipage}\hfill
\caption{Effectiveness$_{\rm e}$ as the function of $r_{\rm zigzag}$ in {\it a)} two and  {\it b)} three dimensions with and without delay.}
\label{fig:zigg}
\end{figure}

The zigzagging has another  parameter: $r_{\rm zigzag}$, which is the critical distance at which the escaper turns off the zigzagging and gets back to the direct escaping to avoid dangerous situations due to the closeness of the chaser(s). Figure \ref{fig:zigg} implies that in 2D $r_{\rm zigzag} = 20$ m is a universal optimum, while, in 3D if $r_{\rm zigzag} > 20$ m, the chasers cannot catch the escaper. With other words i) zigzagging too close to the chasers is disadvantageous in both two and three dimensions, ii) in two dimensions there is an advantageous regime with an optimal parameter for the zigzagging, iii) and in three dimensions the Effectiveness$_{\rm e}$ gets saturated at $r_{\rm zigzag} = r_{\rm sens}$ which means that the evader successfully escape even without the zigzag tactics.


\section{Conclusion }

Based on the corresponding ethological literature we built up a bio-inspired realistic agent-based approach to model collective chasing in continuous-space and discrete-time within closed boundaries for the case when the escaper is significantly faster than the chasers, a situation in which the chasers have no chance to catch the prey alone or without collaboration. Examining the model we have found the following:
\begin{itemize}
\item An optimal group of chasers, which can catch a faster prey, exists in both two and three dimensions when there is a soft, repulsive interaction force between them. This would be impossible without the interaction between the chasers. The optimal group size found here is also comparable to the ones observed in nature when using realistic model parameters.
\item Emergent behaviour occurs - with certain parameters the chasers have the chance to encircle (encage) their prey. These patterns emerge right from the implemented chasing rules  and reflect similar phenomena observed in real biological systems.
\item If the chasers are using the prediction method to forecast their targets' position, their effectiveness increases. 
\item Great delay can completely suppress the chasers' success, but with a long enough prediction time, this can be overcome. 
\item We demonstrated that the zigzag pattern of motion of the erratic escapers can be advantageous, especially when there is delay in the system.
\item Chasing in three dimensions is a much more difficult task for chasers, therefore, the
evader is more likely to survive. 
\item Both the chasers' and the escapers' optimal set of parameters seems to be robust against external noise.
\end{itemize}


\section{Outlook}

Even though our model overcomes those previously proposed in many aspects (e.g. two and three dimensions, delay, erratic escaper, prediction), it lets many directions of the topic to be studied and extended, like i;) many chasers vs. many escapers is just as interesting or even more escapers vs. a small group of chasers ii;) applying evolutionary optimization methods on the species iii;) finding real-life applications iv;) equipping agents with machine learning or real-time adaptive algorithms.

\section{Acknowledgments}

This work was partly supported by the following grants: J\'anos Bolyai Research Scholarship of the Hungarian Academy of Sciences (BO/00219/15/6); K\_16 Research Grant of the Hungarian Academy of Sciences (K 119467).


\end{document}